\documentclass[12pt,onecolumn]{IEEEtran}

\usepackage{amsmath, amssymb, amsthm}

\def\F{{\mathbb F}}
\def\Z{{\mathbb Z}}
\def\Q{{\mathbb Q}}
\newtheorem{thm}{Theorem}[section]
\newtheorem{prop}[thm]{Proposition}

\newtheorem{lem}[thm]{Lemma}
\newtheorem{cor}[thm]{Corollary}
\theoremstyle{definition}

\newtheorem{rem}[thm]{Remark}
\newtheorem{ex}[thm]{Example}
\newenvironment{preuve}[1][Proof]{\noindent\textbf{#1.} }{\ \rule{0.5em}{0.5em}}

\begin{document}

%opening
\title{A new class of codes for Boolean masking of cryptographic computations }

\author{Claude Carlet \hspace{1.3cm} Philippe Gaborit \hspace{1.3cm} Jon-Lark
Kim \hspace{1.3cm} Patrick Sol\'{e}\thanks{CC is with LAGA, Universities of Paris 8
and Paris 13; CNRS, UMR 7539, University of Paris 8, Department of Mathematics,
 2 rue de la libert\'e, 93526 Saint-Denis cedex 02, France. PG is with XLIM-DMI, Universit\'e de Limoges,
123  Avenue Albert Thomas, 87060 Limoges Cedex, France. JLK is with  Department of Mathematics, University of Louisville,
 Louisville, KY 40292, USA and Department of Mathematics, POSTECH, Pohang, Gyungbuk 790-784, South Korea. PS is with CNRS/LTCI, UMR 5141, Telecom ParisTech
46 rue Barrault 75 634 Paris cedex 13, France and MECAA, Math Dept of King Abdulaziz University, Jeddah, Saudi Arabia. }}

\date{}

\maketitle

\begin{abstract}
We introduce a new class of rate one-half binary codes: {\bf complementary information set codes.}
A binary linear code of length $2n$ and dimension $n$ is called a
complementary information set code (CIS code for short) if it has
two disjoint information sets. This class of codes contains self-dual codes as a
subclass. It is connected to graph correlation immune Boolean functions of
use in the security of hardware implementations of  cryptographic primitives.
Such codes permit to improve the cost of masking cryptographic algorithms
against side channel attacks.
In this paper we investigate this new class of codes:
we give optimal or best known CIS codes of length $<132.$
We  derive general constructions based on cyclic codes and on double circulant codes. We derive a Varshamov-Gilbert bound for
long CIS codes, and show that they can all be classified in small
lengths $\le 12$ by the building up construction. Some nonlinear permutations
are constructed by using $\Z_4$-codes, based on the notion of dual
distance of an unrestricted code.
\end{abstract}

{\bf Keywords:} cyclic codes, self-dual codes, dual distance, double circulant codes,
$\Z_4$-codes
\section{Introduction}

Since the seminal work of P. Kocher \cite{K96,K99}, it is known that
the physical implementation of cryptosystems on devices such as smart
cards leaks information which can be used in differential power
analysis or in other kinds of side channel attacks. These attacks can
be devastating if proper counter-measures are not included in the
implementation. A kind of counter-measure, which is suitable for both
hardware and software cryptographic implementations and does not rely
on specific hardware properties is the following. It consists in applying a kind of secret sharing
method, changing the variable representation (say $x$) into
randomized shares $m_1,m_2,\dots ,m_{d+1}$ called masks such that $x=m_1+m_2+\dots
+m_{d+1}$ where $+$ is a group operation - in practice, the XOR.
Since the difficulty of performing an attack of order $d$ (involving $d
+1$ shares) increases exponentially with $d$, it was believed until
recently that for increasing the resistance to attacks, new masks have
to be added, thereby increasing the order of the countermeasure, see
\cite{RP}. But in these schemes, the profusion of masks implies a
heavy load on the true random number generator, which significantly
degrades the performance of the device.
Moreover, the solution in \cite{RP} bases itself on a mask refreshing
operation for which no secure implementation has been detailed so far.
Now, it is shown in~\cite{HM:WISTP-11} that another option consists in
encoding some of the masks, which is much less costly than adding
fresh masks. At the order 1, this consists in representing $x$ by the
ordered pair $(F(m),x+m)$,  where $F$ is a well chosen permutation,
rather than $(m,x+m)$.
The encoding can even be free, in the case it can be merged in a
single table with the downstream function.
Notably, it is demonstrated in \cite{leakage,leakage-bis} that the
same effect as adding several masks can be obtained by the encoding of
one single mask. \\

This method, called {\bf leakage squeezing}, uses permutations $F: {\Bbb F}_2^n\rightarrow {\Bbb F}_2^n$, such that, given some integer $d$ as large as possible,
 for  every pair of vectors $a,b\in {\Bbb F}_2^n$ such that the concatenated vector $(a,b)$ is nonzero and has Hamming weight  $<d$, the value of the Walsh-Hadamard transform of $F$ at $(a,b)$,
 is null. We call such functions {\bf $d$-GCI}, for {\bf Graph Correlation Immune} of order $d$ since  the condition is equivalent
to saying that the indicator function of the graph $\{(x,F(x));\, x\in {\Bbb  F}_2^n\}$ is correlation immune of order $d-1$ \cite{Car-chap}. 
Thus a $d$-GCI function is a protection against an attack of order $d-1.$ These functions were introduced in
 \cite{leakage} where it is proved that the existence of such Boolean functions when they are {\bf linear} is equivalent to the existence of
binary linear codes with parameters $[2n, n, \ge d]$ having two disjoint information sets.
Based on this equivalence we say that a binary linear code of length $2n$ and dimension $n$ is
{\bf Complementary Information Set} (CIS for short) with a partition
$L,\, R$ if there is an information set $L$ whose complement $R$ is
also an information set. In general, we will call the partition written in pidgin Maple as
$[1..n],\,[n+1 ..2n]$ the {\bf systematic partition}.
More explicitly, we describe CIS codes with relation to permutations as follows.

Assuming a  systematic possibly nonlinear code $C$ of length $2n$ of the form
$$C=\{(x,F(x) )|\, x \in \F_2^n\},$$
the permutation is constructed as the map $ x \mapsto F(x).$ In that setting $C$ is CIS by definition if and only if $F$ is a bijection.
 %every vector of length $n$ is the left half of a unique codeword and the image by the permutation of this vector is the right half of this codeword.
 When $C$ is a linear code, we can also consider a systematic generator matrix $(I, A)$ of the code,
where $I$ is the identity matrix of order $n$ and $A$ is a square matrix of order $n$. Then $F(x)=xA,$ and the CIS condition reduces to the fact that $A$ is nonsingular.

On the other hand, since the
complement of an information set of a $[2n,n]$ linear code is an information set for its dual
code, it is clear that systematic {\bf self-dual codes}  are CIS with the
systematic partition. It is also clear that the dual of a CIS code
is CIS. Hence CIS codes are a natural generalization of  self-dual
codes.

In the present work we give explicit constructions of  optimal CIS
codes of length $<132,$ and derive a Varshamov-Gilbert bound for
long CIS codes\footnote{While in cryptographic practice $n$ is not so large, the coding problem is of sufficient intrinsic interest to motivate the study of long codes.}. We give general constructions based on cyclic codes and double circulant codes.
 We show that all CIS codes of length $\le 12$ can be classified by the
building up construction, an important classification tool for self-dual codes \cite{Kim01}. We go back to the notion of graph
correlation immune function, and based on the notion of dual
distance we give a class of unrestricted codes giving Boolean
functions with immunity of higher degree than the functions in the
same number of variables obtained from linear CIS codes (the practical performance of non linear bijections is explored in the companion paper \cite{leakage-bis} ). In
particular the best codes we obtain in that way are binary images of
certain $\Z_4$-codes, the quadratic residue codes \cite{BSC}. Notice moreover that
the notion of being CIS is not trivial since there exists at least one optimal code (a
$[34,17,8]$ code)  which is not CIS.

The material is organized as follows. Section 2 collects the necessary notation and definitions. Section 3 extends the notion of CIS codes
 to possibly nonlinear codes. Section 4 contains the general constructions of CIS codes using cyclic codes and double circulant codes. Section 5 describes the known optimal linear CIS codes of length $<132.$
Section 6 is dedicated to the building up construction and gives a classification of CIS codes of length $\le 12$. Section 7 derives a VG bound on long linear CIS codes without relying on previous knowledge on
the asymptotic performance of self-dual codes. All our computations were done using Magma~\cite{magma}.

\section{Notation and definitions}

\subsection{Binary codes}
Let $C$ be a binary linear code. Its parameters are formatted as $[n,k,d]$ denoting length, dimension, and minimum distance. By an unrestricted code we shall mean a possibly nonlinear code.
 The dual $C^{\perp}$ of a linear code $C$ is understood to be with respect to the standard inner product.
The code $C$ is {\bf self-dual} if $C=C^{\perp}$ and  {\bf isodual} if $C$ is equivalent to $C^{\perp}.$
A self-dual code is {\bf Type II} if the weight of each codeword is a multiple of four and {\bf Type I} otherwise.
A binary (unrestricted) code $C$ of length $n$ is called {\bf systematic} if there  exists a  subset $I$ of $\{1,\cdots,n\}$ called an
{\bf information set} of $C$, such that every possible tuple of length $|I|$ occurs in exactly one codeword within the specified coordinates $x_i;\; i\in I$.
 We call {\bf CIS (unrestricted) code} a systematic code which admits two complementary information sets.
The {\bf Hamming weight} $w(z)$ of a binary vector $z$ is the number of its nonzero entries.
The {\bf weight enumerator $W_C(x,y)$} of a code $C$  is the homogeneous polynomial defined by
$$W_C(x,y)=\sum_{c\in C} x^{n-w(c)}y^{w(c)}.$$
The code $C$ is {\bf formally self-dual} or fsd for short, if its weight enumerator is invariant under the MacWilliams transform, that is,
$$W_C(x,y)=W_C(\frac{x+y}{\sqrt{2}}, \frac{x-y}{\sqrt{2}}). $$
It is furthermore called {\bf even} if $$W_C(x,-y)=W_C(x,y)  ,$$ and {\bf odd} otherwise.
The generator matrix of a $[2n,n]$ code is said to be in {\bf systematic form} if it is blocked as $(I,A)$ with $I$ the identity matrix of order $n.$
If $A$ is circulant then $C$ is said to be {\bf double circulant.}
%In particular, a linear code is  { systematic} if its generator matrix can be put { systematic form}.

\subsection{Boolean functions}
A permutation $F$ of $\F_2^n$ is any bijective map from  ${\Bbb F}_2^n\rightarrow {\Bbb F}_2^n.$ Its Walsh-Hadamard transform at $(a,b)$ is defined as
$$ \widehat{F}(a,b)=\sum_{x\in {\Bbb F}_2^n}(-1)^{ a\cdot x + b\cdot F(x)},$$ where $a\cdot x$ denotes the scalar product of vectors $a$ and $x$.
\subsection{Dual distance}
If $C$ is a binary code, let $B_i$ denote its distance distribution,
that is,
$$B_i=\frac{1}{|C|}|\{(x,y)\in C\times C \, |\, d(x,y)=i\}|$$
The dual distance distribution $B_i^{\perp}$ is the MacWilliams transform of the distance distribution, in the sense that
$$D_C^{\perp}(x,y)=\frac{1}{|C|}D_C(x+y,x-y),$$
where $$D_C(x,y)=\sum_{i = 0}^nB_i x^{n-i}y^i, $$ and $$D_C^\perp (x,y)=\sum_{i = 0}^nB_i^\perp x^{n-i}y^i.$$ The {\bf dual distance} of $C$ is
the smallest $i>0$ such that $B_i^{\perp}\neq 0.$ When $C$ is
linear, it is the minimum distance of $C^{\perp}$, since $D_C^\perp (x,y)=D_{C^\perp}(x,y)$.

For $u \in \F_2^n,$ define the {\bf character} attached to $u$ evaluated in $C$ as
$$ \chi_u(C)=\sum _{v\in C}(-1)^{u \cdot v}.$$
\subsection{$\Z_4$-codes}
Recall that the Gray map $\phi$ from $\Z_4$ to ${\Bbb F}_2^2$ is defined by \\
$$ \phi(0)=00, \phi(1)=01,\phi(2)=11,\phi(3)=10.$$
This map is extended componentwise from $\Z_4^n$ to ${\Bbb
F}_2^{2n}.$ A $\Z_4$-code of length $n$ is a $\Z_4$-submodule of
$\Z_4^n.$ The {\bf binary image} $\phi(C)$ of a $\Z_4$-code $C$ is
just $\{\phi(c)|\,c\in C\}.$ In general a $\Z_4$-code $C$ is of type
$4^k2^l$ if $C\approx \Z_4^k \Z_2^l$ as additive groups. A $\Z_4$-code is called {\bf free} if $l=0.$ An important class of $\Z_4$-codes is that of $QR(p+1)$ where QR stands for {\bf  Quadratic Residue codes} and $p$ is a prime
congruent to $\pm 1 $ modulo $8.$ They were introduced as extended cyclic codes in \cite{BSC}, based on Hensel lifting of classical binary quadratic residue codes \cite{MS77}.

%%%%%%%%%%%%%%%%%%%%%%%%%%%%%%%%%%%%%%%%%%%%%%%%%%%%%%%%%%%%%%%%%%
\section{Generalization to non-linear codes}

Recall that in the context of \cite{leakage,leakage-bis}
we need to use Boolean permutations -  more precisely, permutations $F: {\Bbb F}_2^n\rightarrow {\Bbb F}_2^n$, such that, given some integer $d$ as large as possible,
 for  every pair of vectors $a,b\in {\Bbb F}_2^n$ such that the concatenated $(a,b)$ is nonzero and has Hamming weight  $<d$, the value of the Walsh-Hadamard transform of $F$ at $(a,b)$,
 is null. Such Boolean functions are called {\bf $d$-GCI}, for {\bf Graph Correlation Immune.}

In \cite{leakage-bis} the following proposition is proven:
\begin{prop}\label{them}
The existence of a linear $d$-GCI function of $n$ variables  is
equivalent to the existence of a CIS code of parameters $[2n,n,\ge
d]$ with the systematic partition.
\end{prop}
To generalize this result to systematic, possibly nonlinear codes,
we attach to such a vectorial function $F$ the code
$C_F$ defined by the relationship
$$C_F=\{ (x,F(x)) | \, x \in {\Bbb F}_2^n\}.$$
\begin{thm}\label{me}
The permutation $F: {\Bbb F}_2^n\rightarrow {\Bbb F}_2^n$ is  a
$d$-GCI function of $n$ variables if and only if the code $C_F$ has dual
distance $\ge d.$
\end{thm}

\begin{preuve}
The proof follows immediately by the characterization of the dual distance of a code $C$  in terms of characters $\chi_{u}(C)$ of $C$ regarded as an element
 in the group algebra $\Q[\F_2]$ \cite[Chap. 5, Theorem 7]{MS77}. Essentially, this characterization says that $d^{\perp}$ is the smallest non zero weight
 of a vector $u\in \F_2^n$ such that $\chi_{u}(C)\neq 0.$ Note that the
value of the Walsh-Hadamard transform of $F$ at $(a,b)$ is  $\chi_{u}(C)$ for $u=(a,b)$ and $C=C_F.$
\end{preuve}

Since the dual of a CIS code is also CIS, Proposition \ref{them}
follows at once from Theorem \ref{me}. This theorem also allows us
to construct better GCI functions by using nonlinear codes.
 Define a free $\Z_4$-code of length $n$ with $2^n$ codewords to be {\bf CIS} if it contains two disjoint information sets.
\begin{thm} \label{thm:z4-CIS}
%The permutation $F: {\Bbb F}_2^n\mapsto {\Bbb F}_2^n$ is  a $d-$GCI function of $n$ variables if and only if the code $C_F$ has dual distance $>d.$
If $\cal C$ is a free systematic $\Z_4$-code of length $n$ with
$2^n$ codewords, then its binary image is a systematic code of the
form $C_F$ for some $F.$ Furthermore, $\cal C$ is CIS with systematic partition if and only if $F$
is one-to-one.
\end{thm}
\begin{preuve}
By hypothesis the generator matrix of $\cal C$ is of the shape $(I,A)$ and therefore any codeword of $\cal C$ can be cast as $(x,{\cal F}(x))$ for some $\Z_4$ linear map $\cal F$ from $\Z_4^n$ to $\Z_4^n.$ The result follows by taking
the binary image of each codeword. The CIS property entails that ${\cal F}$ is bijective, and, consequently, so is $F.$ Conversely, if $F$ is bijective then $[n+1..2n]$ is an information set.
\end{preuve}

\begin{ex}
 Consider the Nordstrom Robinson code in length $16.$
 It is the image of the octacode \cite{H+}, which is free and CIS since it is self-dual. It therefore satisfies the hypothesis of Theorem~\ref{thm:z4-CIS} and therefore  can be attached to a $6$-GCI function in $8$ variables, when the best linear CIS code only gives a
$5$-GCI function as shown in the next section.
\end{ex}

\begin{ex}
 Consider $QR32$ a self-dual extended cyclic $\Z_4$-code  \cite{BSBM}.
Its binary image of length $64$ has distance $14,$ which is better
than the best known $[64,32]$ binary code of distance $12.$
Similarly, $QR48$ has a binary image of distance $18$  \cite{BSBM},
when the best binary rate one-half code of length $96$ has distance
$16.$
\end{ex}

\begin{ex}
Recently, Kiermaier and Wassermann~\cite{KieWas} have computed
the Lee weight enumerator of the type II $\Z_4$-code $QR80$ and its
minimum Lee weight $d_L=26$. Hence its binary image has distance
$26$, which is better than the best known $[160,80]$ binary code of
distance $24.$
\end{ex}
\section{General constructions for CIS codes}
We now consider general constructions for the class of CIS codes.
The following two lemmas are, respectively, immediate and well-known.
\begin{lem}\label{systematic}
If a $[2n,n]$ code $C$ has generator matrix $(I,A)$ with $A$ invertible then $C$ is CIS with the systematic partition. Conversely, every CIS code is equivalent to a code with a generator matrix in that form.
\end{lem}
In particular this lemma applies to systematic self-dual codes whose generator matrix $(I,A)$ satisfies $AA^T=I.$ In the following two results, we identify circulant matrices
with the polynomial in $x$ whose $x-$expansion coincides with the first row of the matrix.
\begin{lem}\label{cir_rank}
Let $f(x)$ be a polynomial over $\mathbb F_2$ of degree less than $n$. Then, $\gcd(f(x), x^n -1)=1$ if and only if the circulant matrix generated by $f(x)$ has $\mathbb F_2$-rank $n$.
\end{lem}

\begin{prop}\label{cor_pure}
The double circulant code whose generator matrix is represented by $(1, f(x))$ satisfying Lemma \ref{cir_rank} is a CIS code.
\end{prop}

\begin{preuve}
Combine Lemma \ref{systematic} with Lemma \ref{cir_rank}.
\end{preuve}

In the other direction, a partial converse to Lemma \ref{systematic} is the following.

\begin{prop}\label{rank}
%The double circulant code whose generator matrix is represented by $(1, f(x))$ satisfying Lemma \ref{cir_rank} is a CIS code.
If a $[2n,n]$ code $C$ has generator matrix $(I,A)$ with $rk(A)<n/2$  then $C$ is {\bf not} CIS .
\end{prop}
\begin{preuve}
%Combine Lemma \ref{systematic} with Lemma \ref{cir_rank}.
Let $L=[1..n],$ and $R=[n+1 ..2n].$ If $I$ is an information set, by the rank hypothesis, then $|I \bigcap R |< n/2$ and, consequently, $|I\bigcap L|>n/2.$
 So two distinct information sets must intersect non trivially  in $L.$
\end{preuve}
\medskip

The next proposition is an immediate but useful observation.
\begin{prop}\label{dual}
If $C$ is a $[2n,n]$ code whose dual has minimum weight $1$ then $C$ is not CIS.
\end{prop}
\begin{preuve}
If the dual of $C$ has minimum weight $1$ then the code $C$ has a zero column
and therefore cannot be CIS.
\end{preuve}

\medskip
The previous proposition permits to show it is possible for an optimal code not to be CIS:

\begin{prop}\label{notCIS}
There exists at least one optimal binary code that is not CIS.
\end{prop}
\begin{preuve}
The $[34,17,8]$ code described in the Magma package $BKLC(GF(2),34,17))$
(best known linear code of length $34$ and dimension $17$) is an optimal code
(minimum weight $8$ is the best possible minimum distance for such a code)
whose dual has minimum distance $1$, and therefore is not CIS.
\end{preuve}

\medskip

A special class of CIS codes is obtained from combinatorial matrices \cite{DKS}. Let $A$ be an integral matrix with $0\,/1$ valued entries. We shall say that $A$ is the adjacency matrix of
a {\bf strongly regular graph} (SRG) of parameters $(n,\kappa, \lambda,\mu)$ if $A$ is symmetric, of order $n,$ verifies  $AJ=JA=\kappa J$ and satisfies
$$ A^2=\kappa I+\lambda A+\mu(J-I-A),$$
where $I,\,J$ are the identity and all-one matrices of order $n.$
Alternatively we shall say that $A$ is the adjacency matrix of
a {\bf doubly regular tournament} (DRT) of parameters $(n,\kappa, \lambda,\mu)$ if $A$ is skew-symmetric, of order $n,$ verifies  $AJ=JA=\kappa J$ and satisfies
$$ A^2=\lambda A+\mu(J-I-A).$$
 In the next result we identify $A$ with its reduction mod $2.$
\begin{prop}\label{asssch}
Let $C$ be the linear binary code of length $2n$ spanned by the rows of $(I,M).$ With the above notation, $C$ is CIS if $A$ is the adjacency matrix of a
\begin{itemize}
\item SRG of odd order with
 $\kappa, \lambda$ both even and $\mu$  odd and if $M=A+I$
\item  DRT of odd order with
 $\kappa,\mu$ odd and  $\lambda$  even and if $M=A$
\item SRG of odd order with $\kappa$ even and
 $\lambda, \mu$ both odd and  if $M=A+J$
\item DRT of odd order with $\kappa$ even and
 $\lambda, \mu$ both odd and  if $M=A+J$

\end{itemize}
\end{prop}

\begin{preuve}
In the first case, reduce the quadratic matrix equation modulo $2$ to obtain $$A^2=J+I+A.$$ If $x\in \F_2^n$ is non trivial in $Ker(A+I),$ then
the above equation written as $A(A+I)=J+I,$ entails $Jx=x$ which implies, by studying the eigenspaces of $J$  that
$x$ is the all one vector $\bf 1.$ But because $\kappa$ is even we know that $A {\bf 1}=0.$ This is a contradiction.
So, $Ker(M)$ is trivial and the result follows by Lemma \ref{systematic}.
The proof in the second case is analogous. In the third case the matrix equation becomes $$A^2=J+I.$$ If $x\in Ker(M)$ then $x\in Ker(M^2),$ but, by the hypotheses on $\kappa$ and $n,$ we see that
 $M^2=A^2+J=I,$ yielding $x=0.$ The proof of the fourth case is analogous to that of the third case.
%Combine Lemma \ref{systematic} with Lemma \ref{cir_rank}.
\end{preuve}

Let $q$ be an odd prime power. Let $Q$ be the $q$ by $q$ matrix with zero diagonal and $q_{ij}=1$ if $j-i$ is a square in $GF(q)$ and zero otherwise.
\begin{cor}
 If $q=8j+5$ then the span of $(I,Q+I)$ is CIS. If $q=8j+3$ then the span of $(I,Q)$ is CIS.
\end{cor}

\begin{preuve}
It is well-known \cite{DKS} that $q=4k+1$ then $Q$ is the adjacency matrix of a SRG with parameters $(q,\frac{q-1}{2},\frac{q-5}{4},\frac{q-1}{4}).$
If $q=4k+3$ then $Q$ is the adjacency matrix of a DRT with parameters $(q,\frac{q-1}{2},\frac{q-3}{4},\frac{q+1}{4}).$ The result follows by Proposition \ref{asssch}.
\end{preuve}

The codes obtained in that way are Quadratic Double Circulant codes \cite{G}.  The third and fourth cases of Prop \ref{asssch} cannot apply to either Paley SRG or DRT since
for these we have $\mu-\lambda=1.$

\begin{ex}
 Let $A$ be the adjacency matrix of an SRG of parameters $(57,24,11,9),$ which exists by Brouwer's database \cite{B}. By the third case of Proposition  \ref{asssch} we get
a CIS code of length $114.$
\end{ex}

Now we look at CIS codes constructed from {\bf cyclic codes}. Denote by $C_i$ the code $C$ shortened at coordinate $i$ and by $\overline{C}$ the extension of $C$ by an overall parity check.

\begin{prop}\label{pg}
Let $C$ be a cyclic binary code of odd length $N,$ and dimension $\frac{N+1}{2}.$ If its generator matrix is in circulant form, both $C_1$ and $C_N$
are CIS with the systematic partition. If, furthermore, the weight of the generator polynomial is odd, then $\overline{C}$ is CIS with the systematic partition.
\end{prop}

\begin{preuve}
Recall that in a cyclic code of dimension $k$, consecutive $k$ indices form an information set. The result follows then  for $C_1$ and  $C_N.$ In the extended case, the generator matrix
of $\overline{C}$ is obtained from that of $C$  by inserting a column of $1$'s in position $\frac{N+3}{2}.$ It consists then of two juxtaposed upper triangular and lower triangular, non singular matrices.
\end{preuve}

\section{ CIS codes with record distances}
In this section, we construct optimal or best-known CIS codes of length $2n \le 130$.
We refer to~\cite{GulOst_04} for the highest known minimum distances of rate one-half codes for lengths up to $48.$
In the rest of the section we list what we know of CIS codes of length $<130.$ All statements referring to best rate one-half codes of lengths $>48$ come from \cite{grassl}.
All statements on existence of self-dual codes are from either \cite{GI} or \cite{GII}. We could have used \cite{HM}.
The command $BKLC(GF(2),n,k)$ from the computer package Magma \cite{magma} means the best known binary linear $[n,k]$ code as per \cite{grassl}.
We put a $*$ as an exponent of a distance if the CIS code is optimal as a rate one-half code.
The table captions are as follows
\begin{itemize}
\item bk= obtained the command $BKLC(GF(2),n,k)$ from Magma.
\item dc=double circulant

\item fsd=formally self-dual

\item id=isodual
\item nsd= not self-dual
\item nfsd= not formally self-dual

\item qdc=quadratic double circulant
\item sc=special construction
\item sd= self-dual

\item sqr=shortened quadratic residue code
\item xqdc=extended quadratic double circulant
\item xqr=extended quadratic residue code

\end{itemize}
.
\subsection{ Lengths $2$ to $32$}
$$
\begin{tabular}{|l|l|l|l|l|l|l|l|l|l|l|l|l|l|l|l|l|l|l|}
 \hline
 $2n  $ & 2&  4  & 6& 8&  10 & 12 & 14&16&18&20 & 22 & 24 & 26 & 28 & 30& 32\\ \hline
$d$&       $2^*$&  $2^*$ & $3^*$& $4^*$&   $4^*$ & $4^*$ & $4^*$& $5^*$& $6^*$& $6^*$& $7^*$&  $8^*$&  $7^*$&  $8^*$& $8^*$& $ 8^*$\\ \hline
code &    dc& dc & $\sim$dc  & sd & dc& sd &  sd&$\sim$dc & $\sim$dc& nfsd & id & sd & fsd & dc  & dc  & sd \\
\hline
\end{tabular}
$$
\begin{itemize}
 \item $2n=2, 4$. Double circulant codes of length $2n=2, 4$ with first rows $1$ or $10$, respectively, are optimal CIS codes with minimum distance $2$. They are in fact self-dual.

\item  $2n=6$. Lemma
\ref{cir_rank} does not imply that double circulant codes of this length with $d=3$ are CIS codes.  However, the equivalent code $\{(100~011), (010~101), (001~111)\}$ can be checked to be an optimal CIS code with $d=3$.

\item $2n=8$. The extended Hamming $[8, 4, 4]$ code is an optimal CIS code.

\item $2n=10$. The double circulant with first row $0111$ in~\cite{GulOst_04} is an optimal CIS code with $d=4$ by Lemma
\ref{cir_rank}.

\item $2n=12$. There exists a self-dual $[12, 6, 4]$ code, which is an optimal CIS code.

\item $2n=14$. There exists a self-dual $[14, 7, 4]$ code, which is an optimal CIS code.

\item $2n=16$. There is a unique optimal $[16, 8, 5]$ code~\cite{GulOst_04}, which is also odd formally self-dual. This is a double circulant code with first row $00010111$. It cannot be a CIS code with systematic partition by Lemma \ref{cir_rank}. We need a new representation of this code. Take $C_{16}$ as $BKLC(GF(2), 16, 8)$ in Magma. We check that $C_{16}$ is also an odd formally self-dual $[16, 8, 5]$ code. The determinant of the right half submatrix of the standard generator matrix of $C_{16}$ is $1$. Hence $C_{16}$ is an optimal CIS code.

\item $2n=18$. There is a unique optimal $[18, 9, 6]$ code~\cite{GulOst_04}, which is also odd formally self-dual. It is described as a double circulant code with first row $001001111$. The corresponding polynomial $x^6 + x^3 + x^2 + x + 1$ is factored as $(x^2+x+1)(x^4 + x^3 + 1)$. As $x^9 +1$ has a factor $x^3 +1=(x+1)(x^2+x+1)$, this code cannot be proved to be CIS by Lemma
\ref{cir_rank}. On the other hand, we take $C_{18}$ as $BKLC(GF(2), 18, 9)$ in Magma, which is also a $[18, 9, 6]$ code. The determinant of the right half submatrix of the standard generator matrix of $C_{18}$ is $1$. Hence $C_{18}$ is an optimal CIS code.

\item $2n=20$. There are $1682$ optimal $[20, 10, 6]$ codes~\cite{GulOst_04}. There are exactly eight formally self-dual codes among them~\cite{GulOst_04}. We obtain the first non-formally self-dual optimal CIS code by taking $BKLC(GF(2), 20, 10)$ in Magma notation
with the systematic partition.

\item $2n=22$. We take $BKLC(GF(2), 20, 10)$ in Magma notation
with the systematic partition. This code is isodual and fsd.

\item $2n=24$. The extended Golay code is CIS as a self-dual code. It is also optimal as a rate one-half code.

\item $2n=26$. The fsd isodual code $C_{26,1}$ in the notation of~\cite{GulOst_04} is CIS with the systematic partition.

\item $2n=28$. The even fsd $BKLC(GF(2), 20, 10)$ is not self-dual but still CIS with the sytematic partition by Lemma \ref{systematic}.

\item $2n=30$. We use a double circulant code with generator matrix $(1,f)$ where $f=x^{10} + x^8 + x^7 + x^5 + x^3 + x + 1,$ an irreducible polynomial. This code is CIS with the systematic partition by Proposition \ref{cor_pure}.

\item $2n=32$. There are five extremal Type II $[32,16,8]$ self-dual codes. They are also optimal as rate one-half codes.

\end{itemize}

\subsection{ Lengths $34$ to $64$}
$$
\begin{tabular}{|l|l|l|l|l|l|l|l|l|l|l|l|l|l|l|l|l|}
 \hline
 $2n  $ &   34  & 36& 38&  40 & 42 & 44& 46& 48& 50 & 52 & 54 & 56 & 58 & 60& 62 & 64\\ \hline
$d$&      $8^*$ & $ 8^*$  &  8& 9& $10^*$&  $ 10^*$ &$ 11^*$ &$ 12^*$& 10& 10& 12& 12&  12& 12 & 12 &  12\\ \hline
code &  sc  & sd & sd  & nsd & fsd& fsd& fsd &sd & sd& sd& bk & sd&  qdc & sd  & sd & sd\\
\hline
\end{tabular}
$$
\begin{itemize}

\item $2n=34$. We have checked that $BKLC(GF(2), 34, 17)$ of distance $8$ is not CIS
with systematic partition.
 There are s-extremal self-dual $[34,17,6]$ codes (see \cite{BG}).
Taking the doubly-even subcode of such a code and adding an element
of the shadow we obtain a $[34,17,8]$ code with generator matrix
$(I,A)$ with $A$ given by

{\tiny
\[ \left( \begin{array}{ccccccccccccccccc}
 1 &1 &1 &1 &0 &1 &0 &0 &1 &1 &1 &1 &0 &1 &1 &0 &1 \\
 1  &1 &0 &0 &1 &0 &1 &1 &1 &1 &1 &1 &0 &1 &1 &0 &1 \\
 1 & 0 &1 &0 &1 &0 &0 &0 &0 &0 &1 &1 &0 &1 &1 &0 &1 \\
 0 &1 &1 &0 &1 &0 &0 &0 &1 &1 &0 &0 &0 &1& 1 &0 &1 \\
 1 &1 &1 &1 &1 &0 &1 &0 &1 &0 &1 &0 &1 &1 &1 &0 &0 \\
 1 &1 &1 &1 &1 &1 &1 &1 &1 &1 &1 &1 &1 &0 &1 &1 &1 \\
 1 &0 &0 &0 &1 &1 &0 &0 &1 &1 &0 &1 &0 &1 &0 &0& 1 \\
 1 &0 &0 &1 &1 &1 &1 &0 &1 &0 &1 &1 &0 &1 &1 &1 &1 \\
 0 &1 &0 &1 &0 &1 &0 &0 &0 &1 &0 &0 &1 &1 &0 &0 &1 \\
 0 &1 &0 &0 &0 &1 &1 &0 &1 &1 &0 &1 &1 &1 &1 &1 &1 \\
 0 &0 &1 &0 &1 &1 &0 &1 &1 &1 &1 &1 &0 &1 &1 &1 &1 \\
 0 &0 &1 &1 &1 &1 &0 &0 &1 &0 &1 &0 &0 &1 &0 &0 &1 \\
 0 &0 &0 &0 &0 &1 &0 &1 &0 &1 &1 &0 &1 &1 &1 &1& 0 \\
 0 &0 &0 &0 &1 &0 &0 &1 &0 &1 &0 &1 &1 &1 &1 &0 &1 \\
 0 &0 &0 &0 &1 &1 &1 &1 &0 &0 &1 &1 &0 &1 &0 &0 &0 \\
 0 &0 &0 &1 &0 &1 &0 &0 &1 &1 &1 &1 &0 &1 &0 &1 &0 \\
 0 &0 &0 &1 &1 &0 &1 &1 &1 &1 &1 &1 &1 &1 &1 &1 &0
\end{array} \right)\]
}.

Then it can be checked that this code is CIS with  partition
$L=\{14, \dots, 30\}$
 and $R=\{1, \dots, 13, 31, \dots, 34\}.$

\item $2n=36$. There are many self-dual Type I $[36,18,8]$ self-dual codes. They are also optimal as rate one-half codes.

\item $2n=38$. There are many self-dual Type I $[38,19,8]$ self-dual codes. They are the best known rate one-half codes of that length.

\item $2n=40$. There is an odd fsd isodual $BKLC(GF(2), 40, 20)$ of distance $9,$ the best known rate one-half code of that length. It is CIS by computer check of Lemma \ref{systematic}.

\item $2n=42$. There is an even fsd isodual $BKLC(GF(2), 42, 21)$ of distance $10,$ an optimal rate one-half code of that length. It is CIS by computer check of Lemma \ref{systematic}.

\item $2n=44$. There is an odd fsd isodual $BKLC(GF(2), 44, 22)$ of distance $10,$ an optimal rate one-half code of that length. It is CIS by computer check of Lemma \ref{systematic}.

\item $2n=46$. There is an odd fsd isodual $BKLC(GF(2), 46, 23)$ of distance $11,$ an optimal rate one-half code of that length. It is CIS by computer check of Lemma \ref{systematic}.

\item $2n=48$. There is a unique Type II  $[48,24,12]$ code, an optimal rate one-half code of that length.

\item $2n=50$. There are at least $6$  Type I self-dual codes
of distance $10,$ which is best known as per \cite{grassl}.

\item $2n=52$. There are at least $499$
 Type I self-dual codes
of distance $10,$ which is best known as per \cite{grassl}.
\item $2n=54$. The code $BKLC(GF(2),54,27)$ has distance $11.$  Computing a determinant shows that it is CIS with the systematic partition.

\item $2n=56$. There are Type II self-dual codes of distance $12,$ which is the best known distance for rate one-half codes of that length.

\item $2n=58$. The Quadratic Double Circulant attached to the prime $29$ has distance $12$ and is CIS with the systematic partition by determinant computation.

\item $2n=60$. There are at least $15$ Type I self-dual codes of distance $12,$ the best known distance for rate one-half codes of that length.

\item $2n=62$. There are at least $20$ Type I self-dual codes of distance $12,$ the best known distance for rate one-half codes of that length.

\item $2n=64$. There are at least $3270$ Type II self-dual codes of distance $12,$ the best known distance for rate one-half codes of that length.

\end{itemize}

\subsection{ Lengths $66$ to $100$}

$$
\begin{tabular}{|l|l|l|l|l|l|l|l|l|l|l|l|l|l|l|l|l|l|l|}
 \hline
 $2n  $ & 66&  68  & 70 & 72&  74 & 76 & 78& 80& 82& 84& 86 & 88 & 90 & 92 & 94 & 96 & 98 & 100\\ \hline
$d$&  12     & 13   &15 &15 & 14   &14  &15 &16 &14 &15 & 16& 17 & 18 & 16 & 16& 16 & 17& 18\\ \hline
code &  sd  & bk & bk  & bk & xqr& sd & bk &sd & sd & bk  & sd & sqr& bk & sd  & sd  &  sd & bk & sc\\
\hline
\end{tabular}
$$

\begin{itemize}
\item $2n=66$. There are at least $3$ Type I self-dual codes of distance $12,$ the best known distance for rate one-half codes of that length.

\item $2n=68$. The code $BKLC(GF(2),68,34)$ of distance $13$ is CIS with the systematic partition, by determinant computation.

\item $2n=70$. The code $BKLC(GF(2),70,35)$ of distance $15$ is CIS with the systematic partition, by determinant computation.

\item $2n=72$. The code $BKLC(GF(2),72,36)$ of distance $15$ is CIS with the systematic partition, by determinant computation.

\item $2n=74$. Is $BKLC(GF(2),74,37)$ of distance $14$ CIS? The extended quadratic residue code $[74,37,14]$ is CIS with the systematic partition by Proposition \ref{pg}.

\item $2n=76$. There are at least $2$ Type I self-dual codes of distance $14,$ the best known distance for rate one-half codes of that length.

\item $2n=78$. %The code $BKLC(GF(2),78,39)$ of distance $15$ is CIS with the systematic partition, by determinant computation.
A shortened code of quadratic residue code $[79,40,15]$ leads to a $[78,39,14]$ CIS code with the systematic partition, by Proposition \ref{pg}.

\item $2n=80$. There are at least $15$ Type II self-dual codes of distance $16,$ the best known distance for rate one-half codes of that length.

\item $2n=82$. There is at least $1$ Type I self-dual code of distance $14,$ the best known distance for rate one-half codes of that length.

\item $2n=84$. The code $BKLC(GF(2),84,42)$ of distance $15$ is CIS with the systematic partition, by determinant computation.

\item $2n=86$. There is at least $1$ Type I self-dual code of distance $16,$ the best known distance for rate one-half codes of that length.

\item $2n=88$.  A shortened code of the quadratic residue code $[89,45,17]$ leads to a $[88,44,17]$ CIS code with the systematic partition, by Proposition \ref{pg}.
%The code $BKLC(GF(2),88,44)$ of distance $17$ is CIS with the systematic partition, by determinant computation.

\item $2n=90$. The code $BKLC(GF(2),88,44)$ of distance $18$ is CIS with the systematic partition, by determinant computation.

\item $2n=92$. There are at least $1$ Type I self-dual code of distance $16,$ the best known distance for rate one-half codes of that length.

\item $2n=94$.  There is at least $1$ Type I self-dual code of distance $16,$ the best known distance for rate one-half codes of that length \cite{H++}.

\item $2n=96$. There is at least $1$ Type I self-dual code of distance $16,$ the best known distance for rate one-half codes of that length.

\item $2n=98$. The code $BKLC(GF(2),98,49)$ of distance $17$ is CIS with the systematic partition, by determinant computation.

\item $2n=100$. The code  $BKLC(GF(2),100, 50)$ of distance $18$ is CIS with the systematic partition, since it is obtained by puncturing and double shortening
from a Quadratic Residue code of length $103$ \cite{grassl}. The argument is similar to Proposition \ref{pg}.

\end{itemize}
\subsection{ Lengths $102 $ to $130$}
$$
\begin{tabular}{|l|l|l|l|l|l|l|l|l|l|l|l|l|l|l|l|}
 \hline
 $2n  $ & 102&  104  & 106 & 108&  110 & 112 & 114& 116& 118& 120& 122 & 124& 126 & 128 & 130 \\ \hline
$d$&  19     & 20   &19 & 20 & 18   & 19 & 20 &20 &20 &20 & 20& 21 & 21 & 22 & 22\\ \hline
code &  bk  & bk & qdc  & bk & bk& bk & bk& sc & sc&  sd&  qdc & xqdc & bk& bk & sc \\
\hline
\end{tabular}
$$

\begin{itemize}

\item $2n=102$. The code $BKLC(GF(2),102,51)$ of distance $19$ is CIS with the systematic partition, by determinant computation.

\item $2n=104$. The code $BKLC(GF(2),104,52)$ of distance $20$ is CIS with the systematic partition, by determinant computation.

\item $2n=106$. The code $BKLC(GF(2),106, 53)$ of distance $19$ is CIS by computer search.

\item$2n=108$. The code $BKLC(GF(2),108,54)$ of distance $20$ is CIS with the systematic partition, by determinant computation.

\item$2n=110$. The code $BKLC(GF(2),110,55)$ of distance $18$ is CIS with the systematic partition, by determinant computation.

\item$2n=112$. The code $BKLC(GF(2),112,56)$ of distance $19$ is CIS with the systematic partition, by determinant computation.

\item$2n=114$. The code $BKLC(GF(2),114,57)$ of distance $20$ is CIS with the systematic partition, by determinant computation.

\item $2n=116$. The code $BKLC(GF(2),116, 58)$ is not CIS by Prop
\ref{dual},
  since its dual has minimum weight $1$.
A CIS code can be obtained in the following way: consider $g(x)$ the
generator polynomial
of the $[127,71,19]$ BCH code of length 127 with designed distance $19$.
Consider a generator matrix $G_0$ of the code
obtained by shifting the coefficients of $g(x)$ in the usual way.
Consider now the matrices $G_1$ and $G_2$ obtained
by shifting $G_0$ of respectively $1$ and $2$ positions. We now
construct the matrix $G'=G_0+G_1+G_2$.
We then add to $G'$ a last column of $1'$s and we erase the first row to
obtain a matrix $G$.
The matrix $G$ generates a $[128,70,20]$ code $C$. Now if one shortens
$C$ on
the first twelve columns, one obtains a $[116,58,20]$ code which is CIS
with the systematic partition.

\item $2n=118$. The code $BKLC(GF(2),118, 59)$ is not CIS by Prop \ref{dual},
since its dual has minimum weight $1$.
Similarly to the case $2n=116$, a CIS code can be obtained in the following way:
consider $g(x)$ the generator polynomial
of the BCH code of length 127 and designed distance $19$. Consider a generator matrix $G_0$ of the code
obtained by shifting the coefficients of $g(x)$ in the usual way.
Consider now the matrices $G_1$ and $G_2$ obtained
by shifting $G_0$ of respectively $1$ and $2$ positions. We now construct the matrix $G'=G_0+G_1+G_2$.
We then add to $G'$ a last column of $1$ and we erase the first two rows to obtain a matrix $G$.
The matrix $G$ generates a $[128,69,20]$ code $C$. Now if one shortens the first ten columns of $C$,
one obtains a $[118,59,20]$ code which is CIS with the systematic partition.

\item $2n=120$. There is at least $1$ Type II self-dual code of distance $20,$ the best known distance for rate one-half codes of that length.

\item $2n=122$. Is $BKLC(GF(2),122, 61)$ of distance $20$ CIS? The code $QDC(61)$ of distance $20$ is CIS with the systematic partition, by determinant computation.

\item $2n=124$. Is $BKLC(GF(2),124, 62)$ of distance $20$ CIS? The code $$ExtendCode(BorderedDoublyCirculantQRCode(61))$$ of distance $20$
 is CIS with the systematic partition, by determinant computation.

\item $2n=126$. The code $BKLC(GF(2),126,63)$ of distance $21$ is CIS with the systematic partition, by determinant computation.

\item $2n=128$. The code $BKLC(GF(2),128,64)$ of distance $22$ is CIS with the systematic partition, by determinant computation.

\item $2n=130$. The code $BKLC(GF(2),130, 65)$ of distance $22$ is CIS by computer search.

\end{itemize}

\section{Classification} 
\subsection{Number of equivalence classes of CIS codes} \label{subsec:equiv}
Let $n\ge 2$ be an integer and $g_n$ denote the cardinality of $GL(n,2)$ the general linear group of dimension $n$ over $GF(2).$
 It is well-known (see \cite[p.399]{MS77}), that 
$$
 g_n=\prod_{j=0}^{n-1}(2^n-2^j).
$$
%Let $c_n$ denote the number of conjugacy classes in $GL(n,2)$ the general linear group of dimension $n$ over $GF(2).$
\begin{prop}\label{class}
 The number $e_n$ of equivalence classes of CIS codes of dimension $n\ge 2$    is at most $g_n/n!.$
\end{prop}

\begin{preuve}
By Lemma \ref{systematic} every CIS code of dimension $n$ is equivalent to the linear span of $(I,A)$ for some $A\in GL(n,2).$
 But the columns of such an $A$ are pairwise linearly independent, hence pairwise distinct.
Permuting the columns of $A$ leads to equivalent codes.
\end{preuve}

\begin{ex}
There is a unique CIS code in length $2$ namely $R_2$ the repetition code of length $2.$
 For $n=2,$ the $g_2=6$ invertible matrices reduce to three under column permutation: the identity matrix $I$ and the two triangular matrices $T_1=\left(
     \begin{array}{cc}
      1 & 1 \\
      1 & 0 \\
     \end{array} \right),$ and $T_2=\left(
     \begin{array}{cc}
       0 & 1 \\
       1 & 1 \\
     \end{array}\right).$
The generator matrix $(I,I)$ spans the direct sum $R_2\oplus R_2,$ while the two codes spanned by $(I,T_1)$ and $(I,T_2)$ are equivalent to a code $C_3,$ an isodual code
which is not self-dual. Thus $e_2=2< g_2/2!=3.$
\end{ex}
The numbers $g_n/n!$ grow very fast: $3,\,28,\,840,\,83328.$ They count the number of bases of $\F_2^n$ over $\F_2$ \cite[A053601]{OEIS}. The numbers $e_n$ do not grow so fast as can be seen by
looking at Table I.

\subsection{Building up construction}

The building up construction \cite{Kim01} is known for binary self-dual codes. In this section, we extend it to CIS codes. We show that every CIS code can be constructed in this way.

\begin{lem}\label{lem_subt}
Given a $[2n, n]$ CIS code $C$ with generator matrix $(I_n|A)$ where $A$ is an invertible square matrix of
order $n$, we
can obtain a $[2(n-1), n-1]$ CIS code $C'$ with generator matrix
$(I_{n-1}|A')$, where $A'$ is an invertible square matrix of
order $n-1$.
\end{lem}

\begin{preuve}
Let $a_i$ be the $i$th column of $A$ and $r_i$ be the $i$th row of
$A$ for $1 \le i \le n$. We delete the first column $a_1$. Then for
each $i$, let $r_i'$ be the $i$th row obtained from $r_i$ after the
removal of the first coordinate of $r_i$. Since there are $n$ rows
$r_i'$ in the $(n-1)$-dimensional space, there is $j$ such that
$r_j'=\sum_{i \ne j} c_i r_i'$, where $c_i$ is uniquely determined.
Then delete the $j$th row of $(I_n|A)$ and the $j$th column of $(I_n|A)$
to get a $[2(n-1), n-1]$ CIS code $C'$ with generator matrix
$(I_{n-1}|A')$, where $A'$ is invertible. We remark that $A'$ is a square matrix of order
$n-1$ whose rows consist of the $n-1$ $r_i'$s except for $r_j'$ and
whose columns consist of $a_i'$s ($2 \le i \le n$), each of which
is the $i$th column obtained from $a_i$ after the removal of the
$j$th component of $a_i$.
\end{preuve}

\begin{prop}[Building up construction]\label{prop_building}
Suppose that $C$ is a $[2n, n]$ CIS code $C$ with generator matrix
$(I_n|A)$, where $A$ is invertible with $n$ rows $r_1, \dots, r_n$. Then for any
two vectors $x=(x_1, \cdots, x_n)$ and $y=(y_1, \cdots, y_n)^T$ the
following matrix $G_1$ generates a $[2(n+1), n+1]$ CIS code $C_1$ with the systematic partition:

\begin{equation}\label{eq:G_1}
 G_1 =
 \left(
   \begin{array}{c|cccc|c|c}
    1 & 0 & 0 & \cdots & 0 & z_1 & x \\
 \hline
    0 & 1 &  0 & \cdots      & 0 & y_1 & r_1 \\
    0 & 0 & 1    &  \cdots &0 & y_2 & r_2 \\
\vdots &   &  &   \vdots   &   & \vdots    & \vdots    \\
    0 & 0  & 0  & \cdots & 1 & y_n & r_n \\
   \end{array}
 \right),
\end{equation}
where there are multipliers $c_i$'s satisfying $x= \sum_{i=1}^n c_i r_i$ and $z_1=1+ \sum_{i=1}^n c_i y_i$.
\end{prop}

\begin{preuve} It suffices to show that the rows of the right half of $G_1$ are linearly independent.
Suppose $\alpha (z_1| x)+ \beta_1 (y_1|r_1)+ \cdots  +\beta_n (y_n|r_n) =0$.
Then $\alpha x + \sum_{i=1}^n \beta_i r_i =0$. If $\alpha =0$, then $\beta=0$ for all $i$ as it should. If $\alpha =1$, then $x + \sum_{i=1}^n \beta_i r_i =0$. Since $x= \sum_{i=1}^n c_i r_i$ for unique $c_i$'s, we have $\beta_i =c_i$ for all $i$. Thus $0=z_1 + \sum_{i=1}^n c_i y_i=1+ \sum_{i=1}^n c_i y_i + \sum_{i=1}^n c_i y_i=1$, a contradiction. Thus the rows of the right half of $G'$ are linearly independent
\end{preuve}

\begin{ex}
Let us consider a $[6,3,3]$ CIS code $C$ whose generator matrix is given below.
\[
 G =
 \left(
   \begin{array}{ccc|ccc}
    1 & 0 & 0 &  0 & 1 & 1 \\
    0 & 1 & 0 & 1 & 0 & 1 \\
    0 & 0 & 1 & 1 & 1 & 1 \\
   \end{array}
 \right).
\]

In order to apply Proposition~\ref{prop_building}, we take for example $x=(1,1,0)$ and $y=(1,1,0)^T$. Then $c_1=c_2=1, c_3=0$. Hence $z=1$. In fact, we get the extended Hamming $[8,4,4]$
code whose generator matrix is given below.
\[
 G_1 =
 \left(
   \begin{array}{c|ccc|c|ccc}
  1&  0 & 0 & 0 & 1 & 1 & 1 & 0 \\
  \hline
  0&  1 & 0  &0 & 1 &  0 & 1 & 1 \\
  0&  0 & 1 & 0 &1&  1 & 0 & 1 \\
  0&  0 & 0 & 1 &0&  1 & 1 & 1 \\
   \end{array}
 \right).
\]
Furthermore, starting from $G_1$, we can obtain $G$ back by following the proof of Lemma~\ref{lem_subt}.
\end{ex}

\medskip

%In what follows, we show that every CIS code of length $2n$ can be constructed from a CIS code of length $2(n-1)$ using the building up construction.

\begin{prop}\label{prop:converse} Any $[2n, n]$ CIS code $C$ is
equivalent to a $[2n, n]$ CIS code with the systematic partition
which is constructed from a $[2(n-1), n-1]$ CIS code by using
Proposition~\ref{prop_building}.
\end{prop}

\begin{preuve}
Up to permutation equivalence, we may assume that $C$ is a $[2n, n]$
CIS code with systematic generator matrix $G_2=(I_n|A)$, where $A$ is invertible.
For each $i$ ($1 \le i \le n$), we let $a_i$ be the $i$th column of $A$ and
$r_i$ be the $i$th row of $A$. It suffices to show that there exists
a $2(n-1) \times (n-1)$ systematic generator matrix $(I_{n-1}|A')$ with $A'$ being invertible,
whose extension by Proposition~\ref{prop_building} produces the
matrix $(I_n|A)$ back. We know from Lemma~\ref{lem_subt} that there
exists a $[2(n-1), n-1]$ CIS code $C'$ with systematic generator
matrix $(I_{n-1}|A')$, where $A'$ is invertible. By the remark at the end of the proof of
Lemma~\ref{lem_subt}, $A'$ is a square matrix of order $n-1$ whose
rows consist of the $n-1$ $r_i'$~s except for $r_j'$ and whose
columns consist of $a_i'$~s ($2 \le i \le n$), each of which is the
$i$th column obtained from $a_i$ after the removal of the $j$th
component of $a_i$. We denote the first column $a_1$ of $A$ by
$a_1=(a_1^1, a_1^2, \dots,a_1^j, \dots, a_1^n)^T$ and $a_1'$ by the
column from $a_1$ after the removal of the $j$th component of $a_1$.
We choose $y= a_1'$ and $x=r_j'$. Then it follows from
Proposition~\ref{prop_building} that the below generator matrix
generates a $[2n, n]$ CIS code.

\[
 G_3 =
 \left(
   \begin{array}{c|cccc|c|c}
    1 & 0 & 0 & \cdots & 0 & z_1 & r_j' \\
 \hline
    0 & 1 &  0 & \cdots      & 0 & a_1^1 & r_1' \\
    0 & 0 & 1    &  \cdots &0 & a_1^2 & r_2' \\
\vdots &   &  &   \vdots   &   & \vdots    & \vdots    \\
    0 & 0  & 0  & \cdots & 1 & a_1^n & r_n' \\
   \end{array}
 \right),
\]
where $z_1=1 + \sum_{i \ne j} c_i a_1^i$ and $r_j'=\sum_{i \ne j}c_i
r_i'$ for some $c_i$'s.
 We claim that this $z_1$ is equal to the missing component $a_1^j$ of $a_1'$.
Suppose not. Then $a_1^j=\sum_{i \ne j} c_i a_1^i$, that is, $a_1^j
+ \sum_{i \ne j} c_i a_1^i=0$. Then $(a_1^j|r_j') + \sum_{i \ne
j}c_i(a_1^i|r_i') =(0|0)$. This is a contradiction since the set
$\{r_1=(a_1^1|r_1'), \dots, r_j=(a_1^j|r_j'), \dots,
r_n=(a_1^n|r_n') \}$ is linearly independent. Thus $z_1=a_1^j$.
Therefore, the matrix $G_3$ is equivalent to the original matrix
$G_2$ after permuting rows and columns of $G_3$. This completes the
proof.
\end{preuve}

\begin{cor} Let $C$ be a double circulant CIS $[2n, n]$ code whose generator matrix is of the form $G=(I|A)$,
 where $A$ is an invertible circulant matrix whose first row $r_1$ has odd weight. Then the matrix $G_1$ below generates a CIS code with systematic partition.

\[G_1=
 \left(
   \begin{array}{c|c|c|c}
   1 & 0 \dots 0  & \epsilon & 1 \dots 1 \\
     \hline
   0  &    & 1 &    \\
   \vdots  & I & \vdots & A \\
   0  &    & 1 &    \\
   \end{array}
 \right),
\]
where $\epsilon=0$ if $n$ is odd and $\epsilon=1$ otherwise.
\end{cor}

\begin{preuve} This is the extension of $(I|A)$ using Proposition~\ref{prop_building} where $x$ and $y$ are all one vectors, $x$ is the sum of all rows of $A$, and $z=\epsilon$.
\end{preuve}

\medskip

\begin{rem} Proposition~\ref{prop:converse} implies that one can construct all $[2n+2, n+1]$ CIS codes with systematic partition from the set of {\em all} $[2n, n]$ CIS codes with systematic partition, many of which may be equivalent via column permutations. Here is a natural question. Given a complete list of inequivalent $[2n, n]$ CIS codes with systematic partition, do we always construct all $[2n+2, n+1]$ CIS codes with systematic partition using Proposition~\ref{prop_building}? This may be so in most cases but it may not be true for some cases. Below is a reason.

 Let $C_i$ ($i=1,2$) be a $[2n, n]$ CIS code with generator matrix $(I|A_i)$, where $A_i$ is invertible. Suppose $C_1$ is equivalent to $C_2$ under some $2n$-column permutation (possibly interchanging some columns of the left half coordinates of $C_1$ with some columns in the right half). Let ${\bf D}_i$ ($i=1,2$) be the set of all CIS codes built from $(I|A_i)$ by
Proposition~\ref{prop_building}. Then it may not be true that ${\bf D}_1$ is equivalent to ${\bf D}_2$, that is, for any $C_3 \in {\bf D}_1$, there exists $C_4 \in {\bf D}_2$ such that $C_3$ is equivalent to $C_4$ under some $2n+2$-column permutation, and vice versa. This is because the equivalence of $C_1$ and $C_2$ is via a permutation on $2n$ columns but 
Proposition~\ref{prop_building} is concerned about the right half of the $(n+1) \times (2n+2)$ matrix in the Equation~(\ref{eq:G_1}).
Therefore, given a complete list of inequivalent CIS codes of length $2n$ all of which have systematic partitions, the set of the CIS codes of length $2n+2$ constructed from Proposition~\ref{prop_building} does not necessarily give a complete list of inequivalent CIS codes of length $2n+2$ all of which have systematic partitions. In fact, F. Freibert has informed us that a certain set of the 27 inequivalent $[8,4]$ CIS codes with systematic partition produce only 194 $[10,5]$ CIS codes with systematic partition by Proposition~\ref{prop_building}, although there are exactly 195 inequivalent CIS codes of length 10 (See Proposition~\ref{prop:n=10}).
\end{rem}

In what follows, we give a counting formula similar to a mass formula. This is useful in determining whether a list of inequivalent CIS codes is complete. Recall from Sec.~\ref{subsec:equiv} that $g_n$ denotes the cardinality of $GL(n,2)$.

\begin{prop} \label{prop:mass} Let $n \ge 2$.  Let {\bf C} be the set of all $[2n, n]$ CIS codes and let $S_{2n}$ act on {\bf C} as column permutations of the codes in {\bf C}. Let $C_1, \dots, C_s$ be representatives from every equivalence class of ${\bf C}$ under the action of $S_{2n}$. Let ${\bf C}_{sys}$ be the set of all $[2n, n]$ CIS codes with generator matrix $(I_n|A)$  with $A$ invertible. Suppose that each $C_i \in {\bf C}_{sys}$ ($1 \le i \le s$). Then we have

\begin{equation} \label{eq:mass}
g_n = \sum_{j=1}^s |{\mbox{Orb}}_{S_{2n}}(C_j) \cap {\bf C}_{sys}|,
\end{equation}
where ${\mbox{Orb}}_{S_{2n}}(C_j)$ denotes the orbit of $C_j$ under $S_{2n}$.
\end{prop}

\begin{preuve} Let $A \in GL(n, 2)$. Then each $A$ gives a unique $[2n,n]$ CIS code with generator matrix $(I|A)$. Therefore $g_n = |{\bf C}_{sys}|$. We also note that ${\mbox{Orb}}_{S_{2n}}(C_i) \cap {\bf C}_{sys}$ and
${\mbox{Orb}}_{S_{2n}}(C_j) \cap {\bf C}_{sys}$ are disjoint whenever $i \ne j$. Therefore it is enough to show that each CIS code $C_A$ with generator matrix $(I|A)$ belongs to ${\mbox{Orb}}_{S_{2n}}(C_j) \cap {\bf C}_{sys}$ for a unique $j$. Since $\{C_1, \dots, C_s\}$ is a set of representatives of all CIS codes, $C_A$ belongs to a unique orbit ${\mbox{Orb}}_{S_{2n}}(C_j)$ for some $j$. Clearly $C_A$ is in ${\bf C}_{sys}$ by definition. Therefore $C_A$ belongs to a unique ${\mbox{Orb}}_{S_{2n}}(C_j) \cap {\bf C}_{sys}$ as desired.
\end{preuve}

\subsection{Classification of short CIS codes} \label{sub:class_cis}

We classify all CIS codes of lengths up to $12$ up to equivalence. We successively apply Proposition~\ref{prop_building} from the repetition code of length 2 to obtain lists of inequivalent CIS codes of lengths 4, 6, 8, 10, and 12. For lengths up to 10, we have checked directly that these lists satisfy Equation~(\ref{eq:mass}). For length 12, checking directly that our list satisfies Equation~(\ref{eq:mass}) takes too long. So by inventing equivalence classes among matrices of $GF(n,2)$,
F. Freibert~\cite{Fre} has confirmed that our list of inequivalent CIS codes of length 12 is complete.

\medskip

It is easy to see that any CIS code has minimum distance $\ge 2$.

\begin{itemize}

\item $2n=2$. It is clear that there is a unique CIS code of length $2$, the repetition code.

\item $2n=4$. Applying Proposition~\ref{prop_building} to the repetition code of length $2,$ we show that there are exactly two CIS codes of length $4$. Their generator matrices are $(I|A_{2,1})$ and $(I|A_{2,2})$, where
    \[A_{2,1}=\left(
     \begin{array}{cc}
      1 & 0 \\
      0 & 1 \\
     \end{array} \right),
A_{2,2}=T_2=\left(
     \begin{array}{cc}
      0 & 1 \\
      1 & 1 \\
     \end{array} \right)\]

\item $2n=6$. Using $A_{2,1}$, we have exactly six CIS codes of length $6$, one of which is an optimal code of minimum distance $3$. Similarly using $A_{2,2}$, we have exactly five CIS codes of length $6$. But these latter codes are equivalent to some of the former codes. Generator matrices of the form $(I|A_{3,i})$ ($1 \le i \le 6)$ are given below. Only $A_{3,i}$ are shown in order.

  {\small{  \[\left(
     \begin{array}{ccc}
      1 & 0 & 0 \\
      0 & 1 & 0 \\
      0 & 0 & 1 \\
     \end{array} \right),
\left(
     \begin{array}{ccc}
 1 & 0 & 0 \\
 1 & 1 & 0\\
 0 & 0 & 1\\
     \end{array} \right),
     \left(
     \begin{array}{ccc}
 1 & 0 & 0\\
 1 & 1 & 0\\
 1 & 0 & 1\\
     \end{array} \right),
     \left(
     \begin{array}{ccc}
 0 & 1 & 0\\
 1 & 1 & 0\\
 1 & 0 & 1\\
     \end{array} \right),
     \left(
     \begin{array}{ccc}
 1 & 1 & 1\\
 0 & 1 & 0\\
 0 & 0 & 1\\
     \end{array} \right),
      \left(
     \begin{array}{ccc}
 1 & 1 & 1\\
 1 & 1 & 0\\
 1 & 0 & 1\\
\end{array} \right)\]
     }}

    Hence we have the following.

\begin{prop}\label{prop:n=6}  There are exactly six CIS codes of length $6$. Only one code has $d=3$ and the rest have $d=2$.
\end{prop}

\item $2n=8$. From each of $(I|A_{3,i})$ ($1 \le i \le 6)$ we have many CIS codes of length $8$ and thus consider the equivalence among them. We have the following.

\begin{prop}\label{prop:n=8}  There are exactly $27$ CIS codes of length $8$. Only one code has $d=4$, three have $d=3$, and the rest have $d=2$.
\end{prop}

\item $2n=10$. In a similar manner, we show the following.

\begin{prop}\label{prop:n=10}  There are exactly $195$ CIS codes of length $10$. Four codes have $d=4$, thirty five codes have $d=3$, and the rest have $d=2$.
\end{prop}

\item $2n=12$. Furthermore, using the $195$ CIS codes of length $10$, we have classified CIS codes of length $12$.

\begin{prop}\label{prop:n=12} There are exactly $2705$ CIS codes of length $12$. More precisely, exactly $41$ codes have $d=41$, $565$ codes have $d=3$, and the rest have $d=2$.
\end{prop}

\end{itemize}

We summarize our classification in Table~\ref{tab:classify}. The $i$th column of the table ($i=2,3,4$) represents the number of CIS codes with $d=i$ and the parenthesis gives the number of CIS codes in the order of sd, non-sd fsd, and non-fsd. The last column denotes the total number of CIS codes of the corresponding length. The actual generator matrices of lengths $8, 10, 12$ will be posted in~\cite{Kim2011}.

\begin{table}[thb]
 \caption{Classification of all CIS codes of lengths up to 12 in the order of sd, non-sd fsd, and non-fsd}
 \label{tab:classify}
 \begin{center}
\begin{tabular}{l|l|l|l|l}
 \noalign{\hrule height1pt}
 $2n$ & $d=2$ &  $d=3$  & $d=4$ & Total $\#$  \\
 \hline
 2    & 1 (1+0+0) &   &   & 1 \\
 4    & 2 (1+1+0) &   &   & 2 \\
 6    & 5 (1+2+2)  & 1 (0+1+0) &      & 6 \\
 8    &22 (1+9+12) & 4 (0+2+2) & 1 (1+0+0)    & 27 \\
 10   &156 (2+40+114) & 35 (0+9+26)& 4 (0+2+2)   & 195\\
 12   &2099 (2+318+1779) &565 (0+87+478) & 41 (1+7+33) &2705\\
 \noalign{\hrule height1pt}
\end{tabular}
\end{center}
\end{table}

\section{Long CIS codes}
In this section we show that there are long CIS codes satisfying the VG bound for rate one-half codes,that is with relative distance $\ge H^{-1}(1/2)\approx 0.11$
. We do not use the fact that self-dual codes are CIS. There are polynomial-time constructible binary self-dual codes with relative distance $\approx 0.02$
\cite[p.34, Remark 1]{Q}. We begin by a well-known fact \cite[p.399]{MS77}.
\begin{lem}\label{GL}
The number of invertible $n$ by $n$ matrices is $\sim c2^{n^2},$ with $c\approx 0.29.$
 \end{lem}

Denote by $B(n,d)$ the number of invertible matrices $A$ such that $d$ columns or less of $(I,A)$ are linearly dependent. A crude upper bound on this function can be derived as follows.
\begin{lem}\label{bound}
%The number of invertible $n$ by $n$ matrices is $\sim 2^{n^2}.$

The quantity $B(n,d)$ is $\le M(n,d)$ where $$ M(n,d)=\sum_{j=2}^d \sum_{t=1}^{j-1} {n \choose j-t}{n \choose t } t 2^{n(n-1)} .$$
 \end{lem}
\begin{preuve}
Let $j$ be the size of the linearly dependent family of column vectors of $(I,A),$ with $j-t$ columns of  $I$ and $t$ columns of $A.$ Choose one column amongst $t$ to be obtained as linear combination
of $j-1$ others. Neglecting the invertibility of $A$ we have $n-1$ columns of $A$ to choose freely.
\end{preuve}

Denote by $H(x)=-x\log_2 x -(1-x)\log_2 (1-x)$ the binary entropy function \cite[p.308]{MS77}.
\begin{lem} \label{asymp}
The quantity $M(n,d)$ is dominated by $2^{n^2-n} 2^{2n H(\delta)}$ when $d \sim 2 \delta n$ with $0<\delta <1.$
\end{lem}

\begin{preuve}
 We evaluate the inner sum in $M(n,d)$ by the Chu-Vandermonde identity
$${2n\choose j}=\sum_{t=0}^j {n \choose t}{n \choose j-t}.$$
Then, the outer sum $$\sum_{j=0}^d {2n\choose j}$$ is evaluated by
standard entropic estimates for binomials \cite[p.310]{MS77}. Note that $t\le n,$ a sub-exponential quantity.
\end{preuve}

\begin{prop}
 For each $\delta$ such that $H(\delta) <0.5$ there are long CIS codes of relative distance $\delta.$
\end{prop}

\begin{preuve}
 Consider $(I,A)$ as the parity check matrix of the CIS code and combine Lemmas \ref{GL},\ref{asymp},\ref{bound} to ensure that, asymptotically,
$ |GL(n,2)|>>B(n,d)$ showing the existence of a CIS code of distance $>d,$ for $n$ large enough.
\end{preuve}

\section{Conclusion and Open Problems}
In this article, we have introduced a new class of rate one-half binary codes: CIS codes. In length $2n$ these codes are, when in systematic form,
in one-to-one correspondence with linear bijective permutations in $n$ variables. More generally, bijective permutations correspond to a certain class of systematic codes.
The graph correlation immunity of such permutations is exactly the dual distance of the attached code. Free $\Z_4$ codes of rate one-half can produce such codes
by taking binary images.
It would be a worthwhile task to create a database of $\Z_4$ codes in website form on the model of \cite{grassl}. There should be some good rate one-half free $\Z_4$-codes
in the lengths $40$ to $80.$\\

We have studied the minimum distances of linear CIS codes up to length $130$ and we
have found that it is possible to construct CIS codes as good as the best known minimum distance of rate one-half codes,
 and equal to the best possible distance of
these codes up to length $36.$ Using Table~\ref{tab:classify}, we see that the first length when there is an optimal CIS code that is not self-dual is $4,$ and that the first length when an optimal CIS code cannot be self-dual is $6.$
The first length when an optimal CIS code is not a formally self-dual code is $20.$ Thus this new class of codes is richer than both self-dual, and formally self-dual codes.
While invariant theory is not available to study these codes, a mass formula analogue is Prop. VI.9.  We proved that up to length $130$, there exist CIS codes
with the best known parameters and we also proved that some optimal codes may not be CIS
(for instance in length $34$),
it hence asks the question whether it is possible to find parameters for which
CIS codes have a minimum distance strictly lower than the best linear codes?
More generally, does the CIS property entails an upper bound on the minimum distance?\\

Finally, it is worth extending the definition of  CIS codes to other fields than $\F_2,$ and also to rings. One motivation might be the growing field of
Generalized Boolean functions, that is with ranges other than $\F_2.$ Another motivation like in \S 3 might be to obtain binary CIS codes by some alphabet changing construction.\\

{\bf Acknowledgement}. The authors wish to thank S. Guilley and H. Maghrebi for useful discussions on the leakage squeezing method. The authors also thank Finley Freibert for the discussion on the computational issues of CIS codes. They thank the anonymous referees for helpful remarks that
greatly improved the presentation of the material. J.-L. Kim was partially supported by the Project Completion Grant (year 2011-2012) at the University of Louisville.

\end{document}